\documentclass[aps,prl,twocolumn,superscriptaddress,showpacs,intlimits,amsmath,amssymb,floats,nofootinbib,amsart]{revtex4-1}
\usepackage{graphicx,epsfig,verbatim,physics}
\usepackage{graphics}

\begin{document}
\title{Flat Energy Bands within Antiphase and Twin Boundaries and at Open Edges in Topological Materials}

\author{Linghua Zhu}
    \thanks{Current address: Department of Physics, Virginia Tech, Blacksburg, Virginia 24061}
    \affiliation{Department of Physics, New Jersey Institute of Technology, Newark, New Jersey 07102, USA\\}
\author{Emil Prodan}
    \affiliation{Department of Physics, Yeshiva University, New York, New York 10016, USA\\}    
\author{Keun Hyuk Ahn}
    \email{kenahn@njit.edu}
    \affiliation{Department of Physics, New Jersey Institute of Technology, Newark, New Jersey 07102, USA\\}


\begin{abstract}
A model for two-dimensional electronic, photonic, and mechanical metamaterial systems is presented, which has flat one-dimensional zero-mode energy bands and stable localized states of a topological origin confined within twin boundaries, antiphase boundaries, and at open edges. Topological origins of these flat bands are analyzed for an electronic system as a specific example, using a two-dimensional extension of the Su-Schrieffer-Heeger Hamiltonian with alternating shift of the chains. It is demonstrated that the slow group velocities of the localized flat band states are sensitively controlled by the distance between the boundaries and the propagation can be guided through designed paths of these boundaries. We also discuss how to realize this model in metamaterials.
\end{abstract}

\maketitle

Flat dispersionless energy bands with infinite effective mass, either in electronic, photonic, or mechanical materials, give unique properties to these materials, and have attracted great attention in the recent past~\cite{Chalker2010PRB,Kim2015dPRL,Weeks2010PRB,Wiersma15Phy}. Prominent examples include localized photons in the Lieb photonic lattice~\cite{Mukherjee15PRL,Vicencio15PRL,Poli20172D}, the Mott phenomena and unconventional superconductivity in twisted bilayer graphenes~\cite{Cao2018Nat,Cao2018Nat2}, and the proposal of zero-group-velocity mechanical metamaterials~\cite{Matlack18NMAT}. 
Another class of materials attracting great attention lately are topological materials, which include topological insulators and semimetals~\cite{Hasan2010RMP,Qi2011RMP,Kane2005PRL,Bernevig2006Sci}, phononic materials~\cite{Prodan2009PRL} and metamaterials~\cite{Huber2016TM,Jotzu2014NatLett,Kane2014NatPhys,Lu2015Sci,Lu2017NatPhys,Matlack18NMAT,Meier2016NatComm,Bliokh2015Sci,Klembt2018Nat,Klembt2017APL}.  
Since widely different systems could share phenomena of the same topological origin, the tight binding electronic Hamiltonian for the Haldane model of graphene~\cite{Haldane1988PRL} has been translated to equations describing topological phenomena in photonic, sonic, and mechanical metamaterials, as well as ultracold fermions~\cite{Lu2017NatPhys,Kai2018arXiv,Jotzu2014NatLett,Vila2017PRB,Pal2017NJP}, for example.

In this Letter, we propose a two-dimensional (2D) model system, for which topologically protected flat energy bands arise within the twin boundaries (TB) or the antiphase boundaries (APB), or at the open edges (OE). 
The flat bands are located in the energy gap between bulk bands, which allows the formation of stable localized states, different from the dispersive edge state bands for the Haldane model~\cite{Haldane1988PRL} or the Kane-Mele model~\cite{Kane2005PRL}. It is also demonstrated that the group velocity is tunable by the distance between the boundaries and the propagation can be guided through zig-zag paths, unlike other lattices with flat bands.
We use an electronic tight-binding Hamiltonian as a specific model, and discuss how the same phenomena could be found in metamaterials.

\begin{figure}
    \centering
    \includegraphics[width=0.85\hsize,clip]{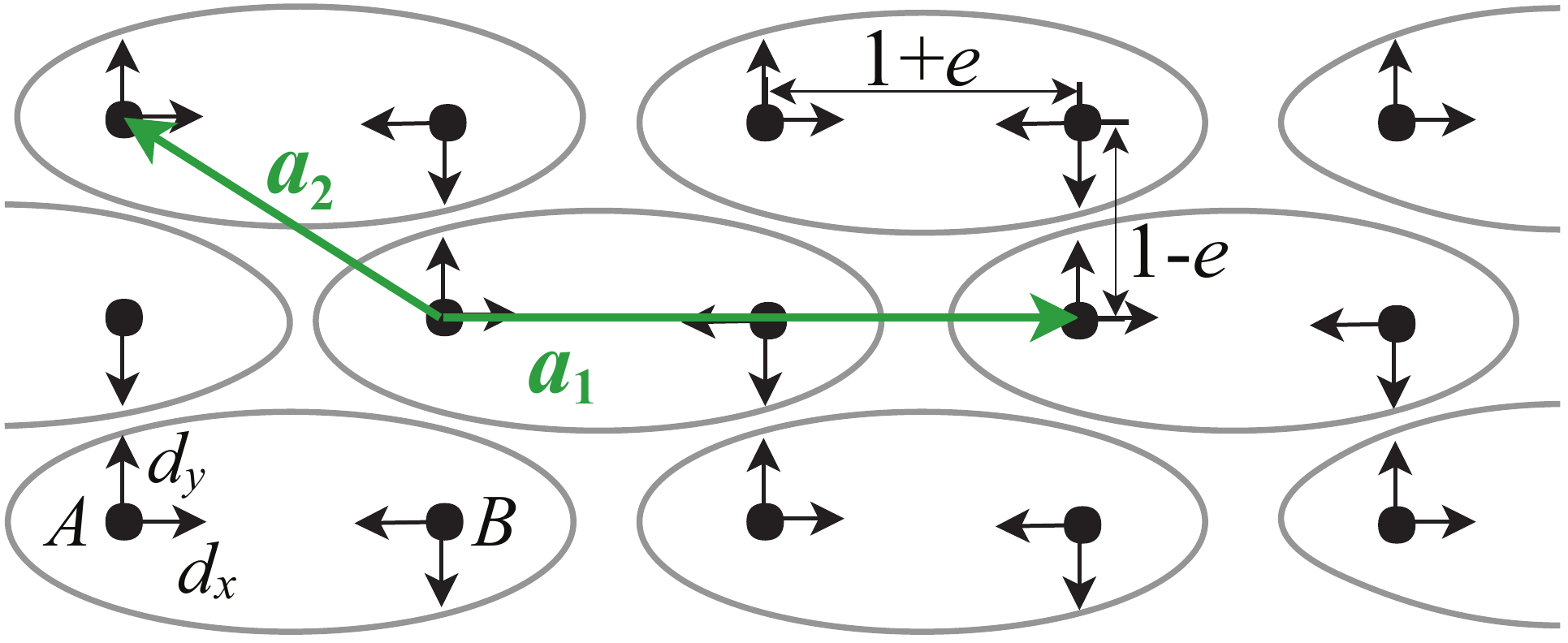}
    \caption{(color online)
    The model system. The black arrows show the staggered distortions parameterized by $d_{x}$ and $d_{y}$, specifically for $A$ atom. Uniform distortions are parameterized by $e$, so that $1+e$ and $1-e$ are the distances between neighboring $A$ and $B$ sites in vertical and horizontal directions respectively, before staggered distortions are introduced. The green arrows represent primitive vectors ${\bf a}_{1}$ and ${\bf a}_{2}$, and the gray ellipses indicate unit cells.
    } \label{fig:model}
\end{figure}

One of the earliest models of the topological insulators is the 1D Su-Schrieffer-Heeger (SSH) model~\cite{ShortCourse,Meier2016NatComm,ChenPRB2014}, for which topologically protected zero energy states could be present at the OE or APB. 
Extension of the SSH model to the 2D space has been studied by simply stacking the 1D SSH chain in the direction perpendicular to the chain~\cite{Delplace11PRB}, which results in dispersive topological edge states. In our study, we extend the 1D SSH model to the 2D space in a different way by shifting every other 1D SSH chain. 
To be specific~\cite{AhnNature2004}, a 2D square lattice is altered first by a uniform square-to-rectangle distortion parameterized by $e$, and then by staggered distortions parameterized by $d_{x}$ and $d_{y}$, shown in Fig.~\ref{fig:model}. 
The nearest-neighbor electron hopping amplitudes depend linearly on interatomic distances, which results in SSH chains in the $x$~$[y]$ direction, shifted and stacked along the $y$~$[x]$ direction with the $inter$chain coupling weaker than the average $intra$chain coupling.
The changes of the phase for the staggered distortions give rise to APB, while the changes of the orientations for uniform rectangular distortions result in TB.
Coherence of underlying lattice structure~\cite{Barsch1984PRL} makes the topological analysis of the TB and APB states possible within convention bulk-boundary correspondence, unlike that of grain boundary states, as recently studied for graphenes and other topological insulators~\cite{Mesaros2010PRB,Takahashi2011PRL,Phillips2015PRB,Slager2016PRB,Rhim2018PRB,FootNote2}.

As shown in Fig.~\ref{fig:model}, a two-atom unit cell is chosen with the unit cell index ${\bf n}=(n_{1},n_{2})$ representing the unit cell at ${\bf{R}}=n_{1}{\bf a}_{1}+n_{2}{\bf a}_{2}$ with primitive vectors ${\bf a}_{1}$ and ${\bf a}_{2}$. Two atoms in the unit cell are labeled as $A$ and $B$. 
The parameters $d_{x}$ and $d_{y}$ specifically represent the staggered components of the distortion of $A$ atom.
The primitive vectors and unit cells are chosen so that one of the primitive vectors is parallel to the boundary and unit cells are not cut through by the boundaries~\cite{Delplace11PRB}.
By considering one spinless electron state at each site, we obtain the following tight binding Hamiltonian for the lattice without APB/TB/OE,
\begin{eqnarray}
   H = &-&(1-e+2d_x)\sigma_+\otimes I-(1-e-2d_x) \sigma_- \otimes S_1   \nonumber \\
  &-&(1+e+2d_y) \sigma_+ \otimes S_2 \nonumber \\
  &-& (1+e-2d_y) \sigma_- \otimes S_1S_2 + {\rm H.c},    
\label{Eq:Ham2}
\end{eqnarray}
where $\quad S_j=\sum_{\bf n} |{\bf n}\rangle \langle {\bf n} + {\bf e}_j|$, ${\bf e}_{1}=(1,0),~{\bf e}_{2}=(0,1)$, $I=\sum_{\bf n} |{\bf n}\rangle \langle {\bf n}|$, $\sigma_+$ and $\sigma_-$ are the raising and lowering materics of the Pauli matrices, and
  $(1,0)^{\rm T}\otimes |{\bf n} \rangle = |A,{\bf n}\rangle $, $(0,1)^{\rm T}\otimes |{\bf n} \rangle = |B,{\bf n}\rangle $ are electron states at corresponding sites.
The hopping amplitude for undistorted lattice and the linear coefficient of the hopping amplitude versus the interatomic distance are chosen as 1 and -1, respectively.  
In various configurations with TB/APB/OE, the coefficients in Eq.~\eqref{Eq:Ham2} become site-dependent,
\begin{eqnarray}
   1\pm e \pm 2d_{x,y} \rightarrow \sum_{\bf n} t_{\bf n} |{\bf n} \rangle \langle {\bf n} |.
\end{eqnarray}

Either with or without TB/APB/OE, the Hamiltonians possess the chiral symmetry, $(\sigma_3 \otimes I)H(\sigma_3 \otimes I) = -H$. For the spinless electron system, the time-reversal symmetry and the particle-hole symmetry are also present with each symmetry squaring to $+1$, which places the system in the BDI class~\cite{Schnyder2008PRB,Qi2008PRB,Kitaev2009AIP,Ryu2010tNJP}. However, when the Hamiltonian is applied to metamaterials, only chiral symmetry may be present, and such systems would belong to the AIII class, and our discussion here uses only the chiral symmetry.
From the classification table of topological condensed matter systems \cite{Schnyder2008PRB,Qi2008PRB,Kitaev2009AIP,Ryu2010tNJP}, there are no strong BDI topological insulators in 2D, although weak topological insulators with non-trivial bulk-boundary correspondence exist \cite{ProdanBook2016}. The latter are characterized in the bulk by a directional 1D winding number, which depends on the orientation of the chosen boundary. A non-trivial bulk topological invariant prompts the emergence of flat edge bands pinned at zero energy. 

The Fourier transformation of the Hamiltonian for the configurations without TB/APB/OE to $\textbf{k}$-space leads to
\begin{eqnarray}
H=\sum_{k_{1},k_{2}}\left [ h(k_{1},k_{2})\sigma _{-}+ h^{*}(k_{1},k_{2})\sigma _{+}\right ]\otimes |{\bf k} \rangle \langle {\bf k} |,
\end{eqnarray}
where 
${\bf k}=(k_{1}/2\pi){\bf b}_{1}+(k_{2}/2\pi){\bf b}_{2}$ with ${\bf b}_{1}$ and ${\bf b}_{2}$ representing reciprocal lattice vectors, and $h(k_1,k_2)=-(1+e+2d_{x})e^{-ik_{2}}-(1+e-2d_{y})e^{i(k_{1}+k_{2})}-(1-e+2d_{y})-(1-e-2d_{x})e^{ik_{1}}.$
The band structure is given by $\varepsilon _{l{\bf k}}(k_{1},k_{2})=(-1)^{l}\left | h(k_{1},k_{2}) \right |$ with $-\pi<k_{1}\leq \pi, -\pi<k_{2}\leq \pi$, and the band index $l=1,2$. For the lattice with a horizontal [vertical] rectangular distortion and $y$-directional [$x$-directional] staggered distortions, that is, $e>0,~d_{x}=0$, $d_{y} \neq 0$ [$e<0,~d_{x} \neq 0$, $d_{y}=0$], a gap opens between the two bands and the system becomes an insulator for the half-filling.
Topology of the system is characterized by the winding number $\nu $~\cite{Delplace11PRB,Zak1989PRL} defined as 
\begin{eqnarray}
\nu({\rm 135^{\circ}}~{\rm or}~{\rm 0^{\circ}}) =\frac{1}{2\pi i}\int_{0}^{2\pi}dk_{\bot}\frac{\partial }{\partial k_{\bot}}\ln{h(k_{1},k_{2})} ,
\label{Eq:Winding}
\end{eqnarray}
where $k_{\bot}=k_{1}$ and $k_{2}$ for boundaries in $135^{\circ}$ and $0^{\circ}$ directions, respectively.
The calculated winding number $\nu(135^{\circ})$ and $\nu(0^{\circ})$ are shown in Table~\ref{Phase_Table} for four possible equivalent distorted insulating states, which reveals that the winding number depends on the signs of the distortions, and the zero modes would be present for TB/APB/OE separating domains of different winding numbers, as discussed in more detail below.

\begin{table}[]
\centering
\begin{tabular}{ccccc}
\hline
~~${e}$~~   & ~~$d_{x}$~~ & ~~$d_{y}$~~ & ~~$\nu(135^{\circ})$~~ & ~~$\nu(0^{\circ})$  \\ \hline
~~$+$~~         & ~~$0$~~             & ~~$+$~~             & ~~0~~   & ~~ -1     \\
~~$+$~~         & ~~$0$~~             & ~~$-$~~             & ~~1~~   & ~~  1   \\
~~$-$~~         & ~~$+$~~             & ~~$0$~~             & ~~0~~   & ~~  0    \\
~~$-$~~         & ~~$-$~~             & ~~$0$~~             & ~~1~~   & ~~  0   \\ \hline
\end{tabular}
\caption{Winding numbers $\nu({\rm 135^{\circ}})~{\rm and}~\nu({\rm 0^{\circ}})$ for boundaries in $135^{\circ}$ and $0^{\circ}$ directions, respectively. The parameters $e$, $d_{x}$, and $d_{y}$ characterize the lattice distortions, as shown in Fig.~\ref{fig:model}.}
\label{Phase_Table}
\end{table}

\begin{figure}
    \centering
    \includegraphics[width=1.00\hsize,clip]{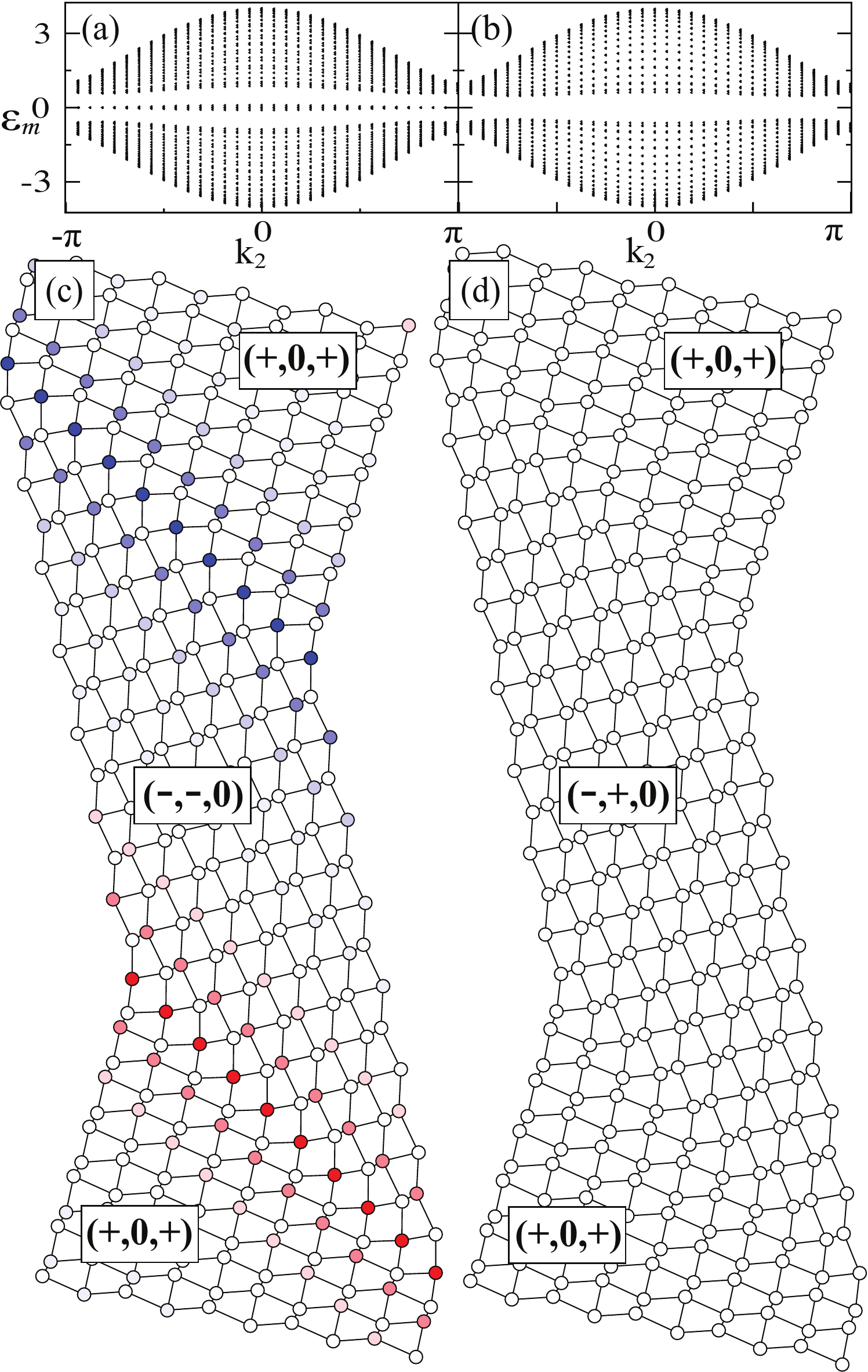}
    \caption{(color online)
    (a), (b) Band structures for the lattice with TB shown (c) and (d), respectively. TB zero-mode bands are present in (a), but absent in (b). (c), (d) Lattices with TB in $135^{\circ}$ direction. Only parts of $32 \times 32$ unit cells with periodic boundary conditions are shown for clarity. The labels in each of the domains represent the signs of $e$, $d_{x}$ and $d_{y}$, which show the difference between (c) and (d) in the sign of $d_{x}$ for the middle domain. Red and blue colors in (c) represent the integrated electron density for the zero-mode band states on $A$ and $B$ sites, respectively.
          } \label{fig:TB} 
\end{figure}

\begin{figure}
    \centering
    \includegraphics[width=1.00\hsize,clip]{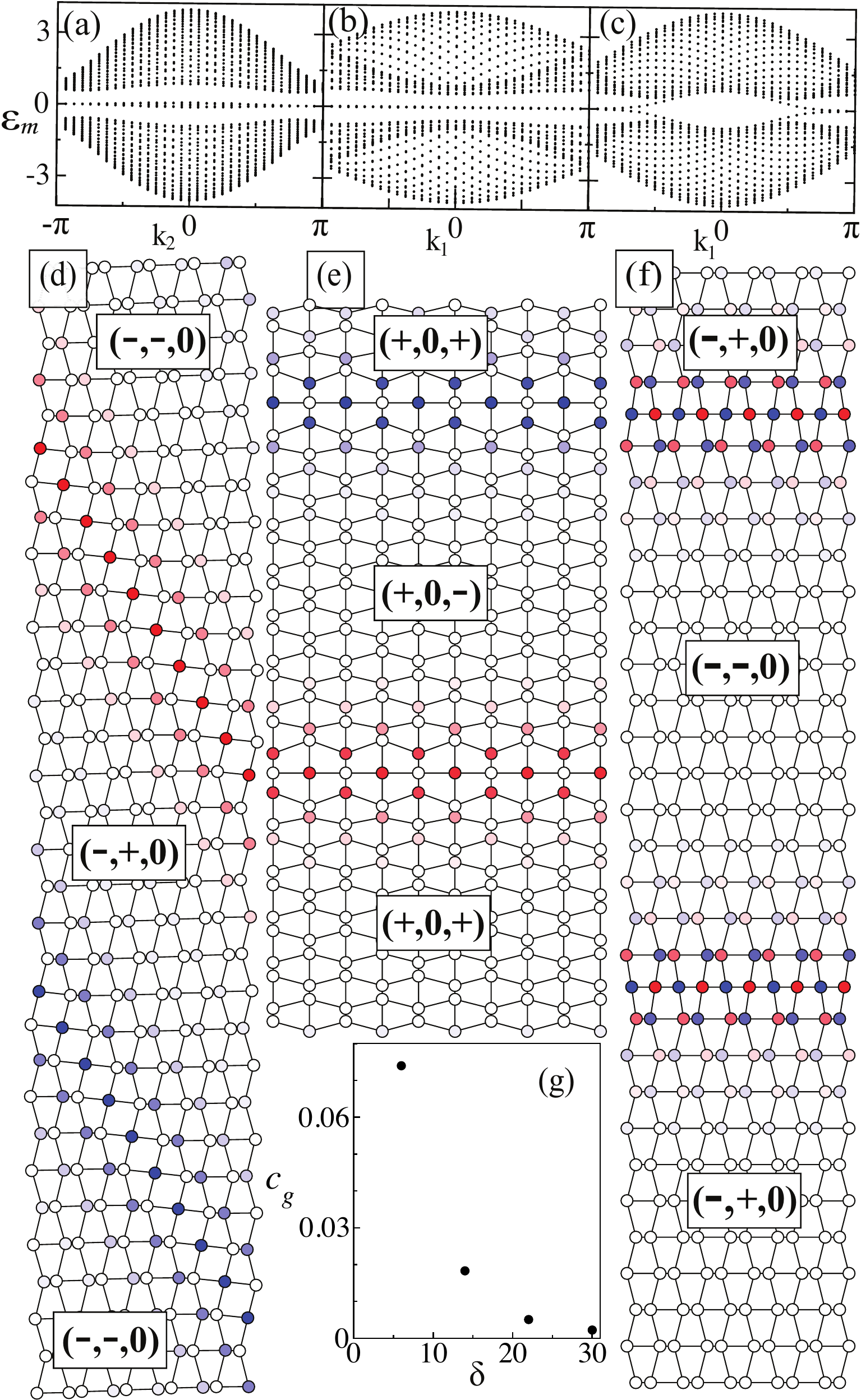}
    \caption{(color online)
    (a), (b), (c) Band structures for the lattices with APB shown in (d), (e) and (f), respectively. APB zero-mode bands are present in (a) and (b), but are absent in (c). Highly dispersive bands inside the gap in (c) are not of topological origin. (d), (e), (f) Lattices with APB in $135^{\circ}$ direction for (d) and $0^{\circ}$ direction for (e) and (f). See the caption for Fig.~\ref{fig:TB}. In (f), the colors represent the integrated electron density for the states within $\varepsilon _{m}=\pm 0.1$. (g) Average group velocity $c_{g}$ versus the number of bonds $\delta$ in the horizontal direction between $135^{\circ}$ direction APB.
          } \label{fig:APB} 
\end{figure}

We present the results obtained by numerical methods for various TB/APB/OE.
The distortion patterns are obtained by relaxing atomic-scale model lattice energy expressions with an anharmonic coupling between uniform and staggered distortions, as described in Supplemental Material~\cite{FootNote1}. Because flat energy bands of zero-modes arise from topological origins, the details of TB/APB/OE configurations do not have much effect on the results presented here. Calculations are carried out for systems of $32 \times 32$ unit cells with periodic boundary conditions and four [two] boundaries in $135^{\circ}$ [$0^{\circ}$] directions. Only parts of the distorted lattices are shown in Figs.~\ref{fig:TB} and \ref{fig:APB} for clarity, with the labels in each domain representing the signs of $e,~d_{x}$, and $d_{y}$.

First, we analyze electronic properties of TB. It is well known that only TB along either $45^{\circ}$ or $135^{\circ}$ direction with respect to the direction of rectangular distortion are stable due to compatibility conditions~\cite{Barsch1984PRL,Shenoy1999PRB}. Figures~\ref{fig:TB}(c) and~\ref{fig:TB}(d) show TB along $135^{\circ}$ direction between domains with horizontal and vertical rectangular distortions.
The difference between the configurations in Figs.~\ref{fig:TB}(c) and ~\ref{fig:TB}(d) is that the staggered distortions, $d_{x}$ in the $e<0$ domain and $d_{y}$ in the $e>0$ domain, have opposite signs in Fig.~\ref{fig:TB}(c) but the same sign in Fig.~\ref{fig:TB}(d).
Therefore, the winding numbers $\nu (135^{\circ})$ for two domains separated by TB are different by one for Fig.~\ref{fig:TB}(c), but identical for Fig.~\ref{fig:TB}(d).
By solving the electronic Hamiltonian numerically, we obtain band structure $\varepsilon _{m}(k_{2})$ with $m=1,\cdots ,64$, shown in Figs.~\ref{fig:TB}(a) and~\ref{fig:TB}(b). 
Zero-mode bands are present in the bulk band gap for Fig.~\ref{fig:TB}(a), while absent for Fig.~\ref{fig:TB}(b). The integrated electron density of the zero-mode states are shown on the lattice in Fig.~\ref{fig:TB}(c) in colors, with red and blue colors indicating \textit{A} and \textit{B} sites, respectively. 
The results show that these states are localized at TB, on the $A$ sites for the lower TB and the $B$ sites for the upper TB. 
The system has one TB state per TB at each $k_{2}$ point, and satisfies the topological bulk-boundary correspondence of $N_{A}-N_{B}=\Delta \nu$, where $\Delta \nu$ is the difference in winding number across the boundary, $N_{A}$ and $N_{B}$ are the number of boundary modes on $A$ and $B$ sites per unit cell. As the distance between TB increases the energies of TB states in Fig.~\ref{fig:TB}(a) approach zero and the zero-mode bands become completely flat throughout the 1D Brillouin zone. 
Such zero-mode flat bands disappear, if the staggered distortion flips the phase in one of the two twin domains, as shown in Figs.~\ref{fig:TB}(b) and~\ref{fig:TB}(d). With a half-filling, TB would be 1D metallic paths~\cite{Kim2016NatComm} in Fig.~\ref{fig:TB}(c), while remain insulating in Fig.~\ref{fig:TB}(d). 

The results for APB are shown in Fig.~\ref{fig:APB}. Unlike the TB, the APB could be formed in any direction. For $135^{\circ}$ (or equivalently $45^{\circ}$) direction APB, Table~\ref{Phase_Table} shows that the winding number $\nu(135^{\circ})$ changes by one whenever the staggered distortion $d_{x}$ or $d_{y}$ changes its phase. This implies zero-mode flat bands are always present at $135^{\circ}$ or $45^{\circ}$ APB, consistent with the numerical results in Figs.~\ref{fig:APB}(a) and \ref{fig:APB}(d). In contrast, the winding number $\nu$ changes by 2 across $0^{\circ}$ [$90^{\circ}$] direction APB for domains with horizontal [vertical] rectangular distortions with $e>0$ [$e<0$], but does not change for domains with vertical [horizontal] rectangular distortions with $e<0$ [$e>0$], consistent with the presence or absence of zero-mode flat bands at the APB in Figs.~\ref{fig:APB}(b), \ref{fig:APB}(c), \ref{fig:APB}(e), and~\ref{fig:APB}(f). Highly dispersive bands inside the gap for the $0^{\circ}$ APB with vertical rectangular distortions in Fig.~\ref{fig:APB}(c) are not of topological origin, and the integrated electron density for states with $\left | \varepsilon _{m} \right |< 0.1$ in Fig.~\ref{fig:APB}(f) shows equal presence on $A$ and $B$ sites at each APB, unlike APB states of topological origin.
It is also found that the presence of zero mode flat bands localized at the OE along $0^{\circ}$ and $135^{\circ}$ directions follows the topological predictions with the winding number outside the OE always zero.

As mentioned at the beginning, the flat bands found for the system could be useful to create stable or slowly moving localized states. While the 2D Lieb lattice provides bands that are flat in the 2D Brillouin zone, our 2D lattice provides bands that are flat in the 1D subspace of the 2D ${\bf k}$-space zone for states localized within TB/APB/OE. 
Such difference gives a unique possibility to our lattice, that is, the tunability of the band dispersion or the group velocity of the localized states by the distance between TB/APB/OE. 
As a demonstration, we consider a pair of $135^{\circ}$ APB, similar to Fig.~\ref{fig:APB}(d) but with a varying number of bonds $\delta $ along the horizontal direction between APB, for $64\times 64$ unit cells, and find that the dispersion of the zero-mode bands increases as $\delta $ decreases. The approximate average group velocity is calculated as $c_{g}=\left [ \varepsilon^{\rm max} _{\rm APB}(k_{2}=0) - \varepsilon^{\rm max} _{\rm APB}(k_{2}=-\pi)\right ]/\pi$, where $\varepsilon^{\rm max} _{\rm APB}(k_{2})$ is the largest among the four zero-mode APB state energies. The result of $c_{g}$ versus $\delta $ is shown in Fig.~\ref{fig:APB}(g), which reveals rapid increase of $c_{g}$ as $\delta $ decreases below around 15. This tunability originates from the hybridization between the states localized around different APBs, like the edge states for the 1D SSH model~\cite{ShortCourse}, and could be useful to design devices with controlled speeds of the propagations for the localized states.    
\begin{figure}
    \centering
    \includegraphics[width=0.97\hsize,clip]{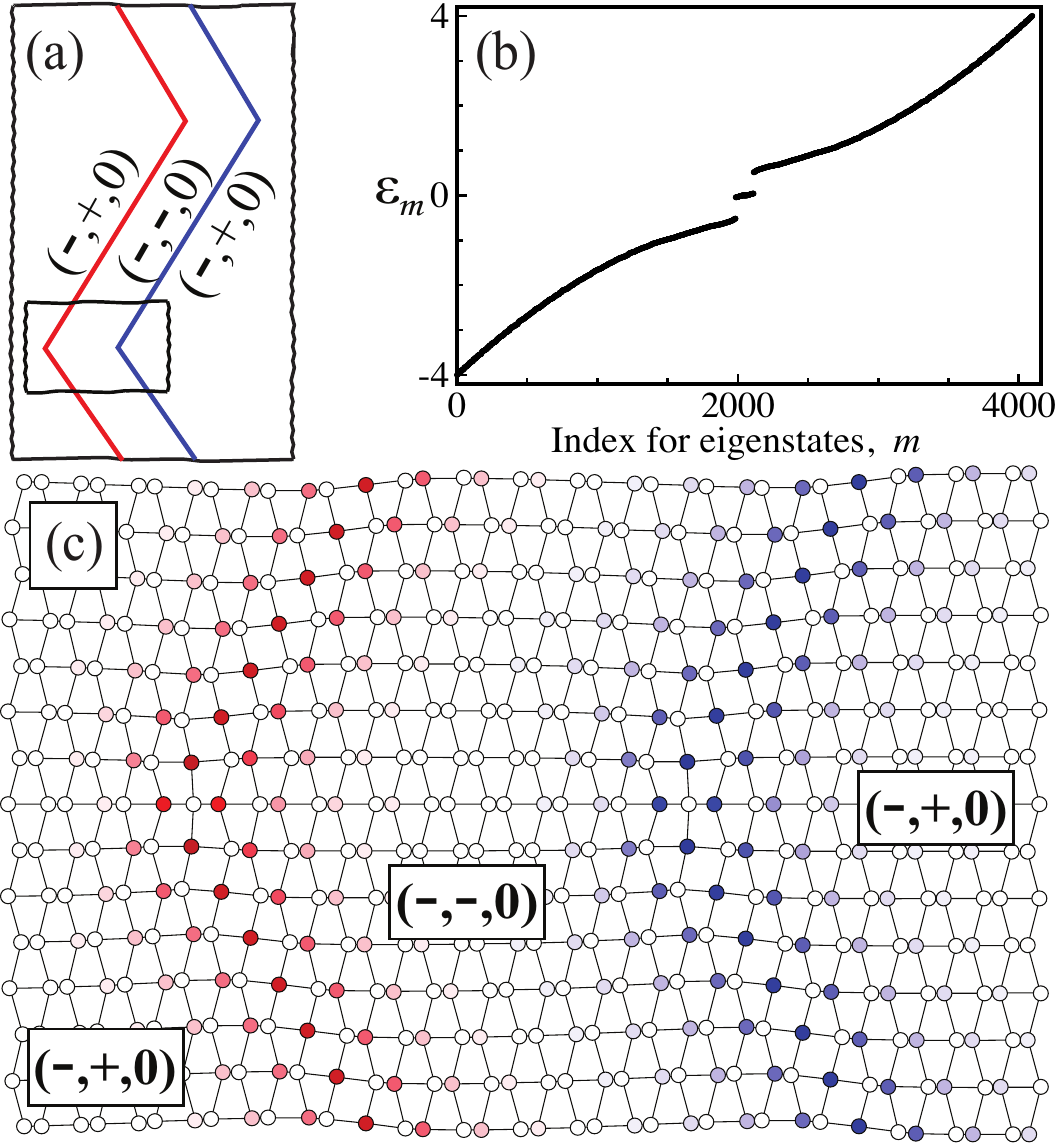}
    \caption{(color online)
     (a) Schematic sketch showing a pair of zig-zag APB with $e<0$ and $d_{y}=0$ for $64 \times 64$ atoms with periodic boundary conditions. From left to right, the phase of $d_{x}$ changes from positive to negative back to positive for the three domains. (b) Plot of energy for eigenstates, $\varepsilon _{m}$, versus the index $m$ for the states in the order of increasing energy. (c) Actual distortion and integrated electron density of zero mode states for the area marked in (a). 
    } \label{fig:Mix}
\end{figure}

To examine whether the propagation of such localized zero-mode states could change their directions without current loss, a pair of zig-zag APB schematically shown in Fig.~\ref{fig:Mix}(a) for $64 \times 64$ atoms are considered, where three domains with a vertical rectangular distortion have the phase of the staggered distortion $d_{x}$ positive, negative and positive from left to right. 
The actual distortion pattern near the kinks in the area marked in Fig.~\ref{fig:Mix}(a) is shown in Fig.~\ref{fig:Mix}(c).
Electronic energy spectrum for the whole distorted lattice is found numerically and energy eigenvalue $\varepsilon_{m}$ versus the index $m$ is displayed in Fig.~\ref{fig:Mix}(b), which shows that the zero-mode APB states are well separated from bulk states in energy in spite of kinks. 
The integrated electron density for these zero-mode states shown in Fig.~\ref{fig:Mix}(c) in colors (see the caption for Fig.~\ref{fig:TB}) indicates that these modes are confined at one sublattice for each zig-zag APB and the current would not be lost at the kinks. Such patterned metamaterials, for example, optical crystals, could be used to guide slowly propagating localized states along designed paths. 

Although we have used electronic Hamiltonian as a specific example, our model can be applied to other systems, such as photonic~\cite{Klembt2018Nat,Klembt2017APL} or mechanical metamaterials or ultracold fermions~\cite{FootNote3}. 
Because chirality symmetry is essential, the metamaterials should have identical on-site energies, or resonances, at all sites, including TB/APB/OE, and the nearest neighbor coupling should have the variations of weakly coupled shifted SSH chains as studied here.
 
In summary, using a 2D model in the weak AIII/BDI topology class with topology-structure coupling, we have demonstrated the presence of flat zero-mode energy bands in the entire 1D Brillouin zone for states localized within TB/APB/OE. It has been found that the flatnesses of these bands and the slow group velocities for the localized zero-mode states could be controlled by the distance between the boundaries and the slow motion of the localized excitations can be guided through a zig-zag path. We propose our model can be realized in various metamaterials, which would open possibilities for unique device applications.

We thank Tsezar F. Seman for the help with figures, and thank members of Academic and Research Computing Systems in the NJIT Information Services and Technology Division for their assistance. The simulations primarily used computational resources managed by NJIT Academic and Research Computing Systems. E.P acknowledges support from the W. M. Keck Foundation.
\bibliography{SSH-Ref}

\begin{thebibliography}{49}%
\makeatletter
\providecommand \@ifxundefined [1]{%
 \@ifx{#1\undefined}
}%
\providecommand \@ifnum [1]{%
 \ifnum #1\expandafter \@firstoftwo
 \else \expandafter \@secondoftwo
 \fi
}%
\providecommand \@ifx [1]{%
 \ifx #1\expandafter \@firstoftwo
 \else \expandafter \@secondoftwo
 \fi
}%
\providecommand \natexlab [1]{#1}%
\providecommand \enquote  [1]{``#1''}%
\providecommand \bibnamefont  [1]{#1}%
\providecommand \bibfnamefont [1]{#1}%
\providecommand \citenamefont [1]{#1}%
\providecommand \href@noop [0]{\@secondoftwo}%
\providecommand \href [0]{\begingroup \@sanitize@url \@href}%
\providecommand \@href[1]{\@@startlink{#1}\@@href}%
\providecommand \@@href[1]{\endgroup#1\@@endlink}%
\providecommand \@sanitize@url [0]{\catcode `\\12\catcode `\$12\catcode
  `\&12\catcode `\#12\catcode `\^12\catcode `\_12\catcode `\%12\relax}%
\providecommand \@@startlink[1]{}%
\providecommand \@@endlink[0]{}%
\providecommand \url  [0]{\begingroup\@sanitize@url \@url }%
\providecommand \@url [1]{\endgroup\@href {#1}{\urlprefix }}%
\providecommand \urlprefix  [0]{URL }%
\providecommand \Eprint [0]{\href }%
\providecommand \doibase [0]{http://dx.doi.org/}%
\providecommand \selectlanguage [0]{\@gobble}%
\providecommand \bibinfo  [0]{\@secondoftwo}%
\providecommand \bibfield  [0]{\@secondoftwo}%
\providecommand \translation [1]{[#1]}%
\providecommand \BibitemOpen [0]{}%
\providecommand \bibitemStop [0]{}%
\providecommand \bibitemNoStop [0]{.\EOS\space}%
\providecommand \EOS [0]{\spacefactor3000\relax}%
\providecommand \BibitemShut  [1]{\csname bibitem#1\endcsname}%
\let\auto@bib@innerbib\@empty
\bibitem [{\citenamefont {Chalker}\ \emph {et~al.}(2010)\citenamefont
  {Chalker}, \citenamefont {Pickles},\ and\ \citenamefont
  {Shukla}}]{Chalker2010PRB}%
  \BibitemOpen
  \bibfield  {author} {\bibinfo {author} {\bibfnamefont {J.~T.}\ \bibnamefont
  {Chalker}}, \bibinfo {author} {\bibfnamefont {T.~S.}\ \bibnamefont
  {Pickles}}, \ and\ \bibinfo {author} {\bibfnamefont {P.}~\bibnamefont
  {Shukla}},\ }\href@noop {} {\bibfield  {journal} {\bibinfo  {journal} {Phys.
  Rev. B}\ }\textbf {\bibinfo {volume} {82}},\ \bibinfo {pages} {104209}
  (\bibinfo {year} {2010})}\BibitemShut {NoStop}%
\bibitem [{\citenamefont {Kim}\ \emph {et~al.}(2015)\citenamefont {Kim},
  \citenamefont {Wieder}, \citenamefont {Kane},\ and\ \citenamefont
  {Rappe}}]{Kim2015dPRL}%
  \BibitemOpen
  \bibfield  {author} {\bibinfo {author} {\bibfnamefont {Y.}~\bibnamefont
  {Kim}}, \bibinfo {author} {\bibfnamefont {B.~J.}\ \bibnamefont {Wieder}},
  \bibinfo {author} {\bibfnamefont {C.~L.}\ \bibnamefont {Kane}}, \ and\
  \bibinfo {author} {\bibfnamefont {A.~M.}\ \bibnamefont {Rappe}},\ }\href@noop
  {} {\bibfield  {journal} {\bibinfo  {journal} {Phys. Rev. Lett.}\ }\textbf
  {\bibinfo {volume} {115}},\ \bibinfo {pages} {036806} (\bibinfo {year}
  {2015})}\BibitemShut {NoStop}%
\bibitem [{\citenamefont {Weeks}\ and\ \citenamefont
  {Franz}(2010)}]{Weeks2010PRB}%
  \BibitemOpen
  \bibfield  {author} {\bibinfo {author} {\bibfnamefont {C.}~\bibnamefont
  {Weeks}}\ and\ \bibinfo {author} {\bibfnamefont {M.}~\bibnamefont {Franz}},\
  }\href@noop {} {\bibfield  {journal} {\bibinfo  {journal} {Phys. Rev. B}\
  }\textbf {\bibinfo {volume} {82}},\ \bibinfo {pages} {085310} (\bibinfo
  {year} {2010})}\BibitemShut {NoStop}%
\bibitem [{\citenamefont {Wiersma}(2015)}]{Wiersma15Phy}%
  \BibitemOpen
  \bibfield  {author} {\bibinfo {author} {\bibfnamefont {D.~S.}\ \bibnamefont
  {Wiersma}},\ }\href@noop {} {\bibfield  {journal} {\bibinfo  {journal}
  {Physics}\ }\textbf {\bibinfo {volume} {8}},\ \bibinfo {pages} {55} (\bibinfo
  {year} {2015})}\BibitemShut {NoStop}%
\bibitem [{\citenamefont {Mukherjee}\ \emph {et~al.}(2015)\citenamefont
  {Mukherjee}, \citenamefont {Spracklen}, \citenamefont {Choudhury},
  \citenamefont {Goldman}, \citenamefont {{\"O}hberg}, \citenamefont
  {Andersson},\ and\ \citenamefont {Thomson}}]{Mukherjee15PRL}%
  \BibitemOpen
  \bibfield  {author} {\bibinfo {author} {\bibfnamefont {S.}~\bibnamefont
  {Mukherjee}}, \bibinfo {author} {\bibfnamefont {A.}~\bibnamefont
  {Spracklen}}, \bibinfo {author} {\bibfnamefont {D.}~\bibnamefont
  {Choudhury}}, \bibinfo {author} {\bibfnamefont {N.}~\bibnamefont {Goldman}},
  \bibinfo {author} {\bibfnamefont {P.}~\bibnamefont {{\"O}hberg}}, \bibinfo
  {author} {\bibfnamefont {E.}~\bibnamefont {Andersson}}, \ and\ \bibinfo
  {author} {\bibfnamefont {R.~R.}\ \bibnamefont {Thomson}},\ }\href@noop {}
  {\bibfield  {journal} {\bibinfo  {journal} {Phys. Rev. Lett.}\ }\textbf
  {\bibinfo {volume} {114}},\ \bibinfo {pages} {245504} (\bibinfo {year}
  {2015})}\BibitemShut {NoStop}%
\bibitem [{\citenamefont {Vicencio}\ \emph {et~al.}(2015)\citenamefont
  {Vicencio}, \citenamefont {Cantillano}, \citenamefont {Morales-Inostroza},
  \citenamefont {Real}, \citenamefont {Mej{\'\i}a-Cort{\'e}s}, \citenamefont
  {Weimann}, \citenamefont {Szameit},\ and\ \citenamefont
  {Molina}}]{Vicencio15PRL}%
  \BibitemOpen
  \bibfield  {author} {\bibinfo {author} {\bibfnamefont {R.~A.}\ \bibnamefont
  {Vicencio}}, \bibinfo {author} {\bibfnamefont {C.}~\bibnamefont
  {Cantillano}}, \bibinfo {author} {\bibfnamefont {L.}~\bibnamefont
  {Morales-Inostroza}}, \bibinfo {author} {\bibfnamefont {B.}~\bibnamefont
  {Real}}, \bibinfo {author} {\bibfnamefont {C.}~\bibnamefont
  {Mej{\'\i}a-Cort{\'e}s}}, \bibinfo {author} {\bibfnamefont {S.}~\bibnamefont
  {Weimann}}, \bibinfo {author} {\bibfnamefont {A.}~\bibnamefont {Szameit}}, \
  and\ \bibinfo {author} {\bibfnamefont {M.~I.}\ \bibnamefont {Molina}},\
  }\href@noop {} {\bibfield  {journal} {\bibinfo  {journal} {Phys. Rev. Lett.}\
  }\textbf {\bibinfo {volume} {114}},\ \bibinfo {pages} {245503} (\bibinfo
  {year} {2015})}\BibitemShut {NoStop}%
\bibitem [{\citenamefont {Poli}\ \emph {et~al.}(2017)\citenamefont {Poli},
  \citenamefont {Schomerus}, \citenamefont {Bellec}, \citenamefont {Kuhl},\
  and\ \citenamefont {Mortessagne}}]{Poli20172D}%
  \BibitemOpen
  \bibfield  {author} {\bibinfo {author} {\bibfnamefont {C.}~\bibnamefont
  {Poli}}, \bibinfo {author} {\bibfnamefont {H.}~\bibnamefont {Schomerus}},
  \bibinfo {author} {\bibfnamefont {M.}~\bibnamefont {Bellec}}, \bibinfo
  {author} {\bibfnamefont {U.}~\bibnamefont {Kuhl}}, \ and\ \bibinfo {author}
  {\bibfnamefont {F.}~\bibnamefont {Mortessagne}},\ }\href@noop {} {\bibfield
  {journal} {\bibinfo  {journal} {2D Mater.}\ }\textbf {\bibinfo {volume}
  {4}},\ \bibinfo {pages} {025008} (\bibinfo {year} {2017})}\BibitemShut
  {NoStop}%
\bibitem [{\citenamefont {Cao}\ \emph {et~al.}(2018{\natexlab{a}})\citenamefont
  {Cao}, \citenamefont {Fatemi}, \citenamefont {Demir}, \citenamefont {Fang},
  \citenamefont {Tomarken}, \citenamefont {Luo}, \citenamefont
  {Sanchez-Yamagishi}, \citenamefont {Watanabe}, \citenamefont {Taniguchi},
  \citenamefont {Kaxiras}, \citenamefont {Ashoori},\ and\ \citenamefont
  {Jarillo-Herrero}}]{Cao2018Nat}%
  \BibitemOpen
  \bibfield  {author} {\bibinfo {author} {\bibfnamefont {Y.}~\bibnamefont
  {Cao}}, \bibinfo {author} {\bibfnamefont {V.}~\bibnamefont {Fatemi}},
  \bibinfo {author} {\bibfnamefont {A.}~\bibnamefont {Demir}}, \bibinfo
  {author} {\bibfnamefont {S.}~\bibnamefont {Fang}}, \bibinfo {author}
  {\bibfnamefont {S.~L.}\ \bibnamefont {Tomarken}}, \bibinfo {author}
  {\bibfnamefont {J.~Y.}\ \bibnamefont {Luo}}, \bibinfo {author} {\bibfnamefont
  {J.~D.}\ \bibnamefont {Sanchez-Yamagishi}}, \bibinfo {author} {\bibfnamefont
  {K.}~\bibnamefont {Watanabe}}, \bibinfo {author} {\bibfnamefont
  {T.}~\bibnamefont {Taniguchi}}, \bibinfo {author} {\bibfnamefont
  {E.}~\bibnamefont {Kaxiras}}, \bibinfo {author} {\bibfnamefont {R.~C.}\
  \bibnamefont {Ashoori}}, \ and\ \bibinfo {author} {\bibfnamefont
  {P.}~\bibnamefont {Jarillo-Herrero}},\ }\href@noop {} {\bibfield  {journal}
  {\bibinfo  {journal} {Nature}\ }\textbf {\bibinfo {volume} {556}},\ \bibinfo
  {pages} {80} (\bibinfo {year} {2018}{\natexlab{a}})}\BibitemShut {NoStop}%
\bibitem [{\citenamefont {Cao}\ \emph {et~al.}(2018{\natexlab{b}})\citenamefont
  {Cao}, \citenamefont {Fatemi}, \citenamefont {Fang}, \citenamefont
  {Watanabe}, \citenamefont {Taniguchi}, \citenamefont {Kaxiras},\ and\
  \citenamefont {Jarillo-Herrero}}]{Cao2018Nat2}%
  \BibitemOpen
  \bibfield  {author} {\bibinfo {author} {\bibfnamefont {Y.}~\bibnamefont
  {Cao}}, \bibinfo {author} {\bibfnamefont {V.}~\bibnamefont {Fatemi}},
  \bibinfo {author} {\bibfnamefont {S.}~\bibnamefont {Fang}}, \bibinfo {author}
  {\bibfnamefont {K.}~\bibnamefont {Watanabe}}, \bibinfo {author}
  {\bibfnamefont {T.}~\bibnamefont {Taniguchi}}, \bibinfo {author}
  {\bibfnamefont {E.}~\bibnamefont {Kaxiras}}, \ and\ \bibinfo {author}
  {\bibfnamefont {P.}~\bibnamefont {Jarillo-Herrero}},\ }\href@noop {}
  {\bibfield  {journal} {\bibinfo  {journal} {Nature}\ }\textbf {\bibinfo
  {volume} {556}},\ \bibinfo {pages} {43} (\bibinfo {year}
  {2018}{\natexlab{b}})}\BibitemShut {NoStop}%
\bibitem [{\citenamefont {Matlack}\ \emph {et~al.}(2018)\citenamefont
  {Matlack}, \citenamefont {Serra-Garcia}, \citenamefont {Palermo},
  \citenamefont {Huber},\ and\ \citenamefont {Daraio}}]{Matlack18NMAT}%
  \BibitemOpen
  \bibfield  {author} {\bibinfo {author} {\bibfnamefont {K.~H.}\ \bibnamefont
  {Matlack}}, \bibinfo {author} {\bibfnamefont {M.}~\bibnamefont
  {Serra-Garcia}}, \bibinfo {author} {\bibfnamefont {A.}~\bibnamefont
  {Palermo}}, \bibinfo {author} {\bibfnamefont {S.~D.}\ \bibnamefont {Huber}},
  \ and\ \bibinfo {author} {\bibfnamefont {C.}~\bibnamefont {Daraio}},\
  }\href@noop {} {\bibfield  {journal} {\bibinfo  {journal} {Nat. Mater.}\
  }\textbf {\bibinfo {volume} {17}},\ \bibinfo {pages} {323} (\bibinfo {year}
  {2018})}\BibitemShut {NoStop}%
\bibitem [{\citenamefont {Hasan}\ and\ \citenamefont
  {Kane}(2010)}]{Hasan2010RMP}%
  \BibitemOpen
  \bibfield  {author} {\bibinfo {author} {\bibfnamefont {M.~Z.}\ \bibnamefont
  {Hasan}}\ and\ \bibinfo {author} {\bibfnamefont {C.~L.}\ \bibnamefont
  {Kane}},\ }\href@noop {} {\bibfield  {journal} {\bibinfo  {journal} {Rev.
  Mod. Phys.}\ }\textbf {\bibinfo {volume} {82}},\ \bibinfo {pages} {3045}
  (\bibinfo {year} {2010})}\BibitemShut {NoStop}%
\bibitem [{\citenamefont {Qi}\ and\ \citenamefont {Zhang}(2011)}]{Qi2011RMP}%
  \BibitemOpen
  \bibfield  {author} {\bibinfo {author} {\bibfnamefont {X.-L.}\ \bibnamefont
  {Qi}}\ and\ \bibinfo {author} {\bibfnamefont {S.-C.}\ \bibnamefont {Zhang}},\
  }\href@noop {} {\bibfield  {journal} {\bibinfo  {journal} {Rev. Mod. Phys.}\
  }\textbf {\bibinfo {volume} {83}},\ \bibinfo {pages} {1057} (\bibinfo {year}
  {2011})}\BibitemShut {NoStop}%
\bibitem [{\citenamefont {Kane}\ and\ \citenamefont
  {Mele}(2005)}]{Kane2005PRL}%
  \BibitemOpen
  \bibfield  {author} {\bibinfo {author} {\bibfnamefont {C.~L.}\ \bibnamefont
  {Kane}}\ and\ \bibinfo {author} {\bibfnamefont {E.~J.}\ \bibnamefont
  {Mele}},\ }\href@noop {} {\bibfield  {journal} {\bibinfo  {journal} {Phys.
  Rev. Lett.}\ }\textbf {\bibinfo {volume} {95}},\ \bibinfo {pages} {226801}
  (\bibinfo {year} {2005})}\BibitemShut {NoStop}%
\bibitem [{\citenamefont {Bernevig}\ \emph {et~al.}(2006)\citenamefont
  {Bernevig}, \citenamefont {Hughes},\ and\ \citenamefont
  {Zhang}}]{Bernevig2006Sci}%
  \BibitemOpen
  \bibfield  {author} {\bibinfo {author} {\bibfnamefont {B.~A.}\ \bibnamefont
  {Bernevig}}, \bibinfo {author} {\bibfnamefont {T.~L.}\ \bibnamefont
  {Hughes}}, \ and\ \bibinfo {author} {\bibfnamefont {S.-C.}\ \bibnamefont
  {Zhang}},\ }\href@noop {} {\bibfield  {journal} {\bibinfo  {journal}
  {Science}\ }\textbf {\bibinfo {volume} {314}},\ \bibinfo {pages} {1757}
  (\bibinfo {year} {2006})}\BibitemShut {NoStop}%
\bibitem [{\citenamefont {Prodan}\ and\ \citenamefont
  {Prodan}(2009)}]{Prodan2009PRL}%
  \BibitemOpen
  \bibfield  {author} {\bibinfo {author} {\bibfnamefont {E.}~\bibnamefont
  {Prodan}}\ and\ \bibinfo {author} {\bibfnamefont {C.}~\bibnamefont
  {Prodan}},\ }\href@noop {} {\bibfield  {journal} {\bibinfo  {journal} {Phys.
  Rev. Lett.}\ }\textbf {\bibinfo {volume} {103}},\ \bibinfo {pages} {248101}
  (\bibinfo {year} {2009})}\BibitemShut {NoStop}%
\bibitem [{\citenamefont {Huber}(2016)}]{Huber2016TM}%
  \BibitemOpen
  \bibfield  {author} {\bibinfo {author} {\bibfnamefont {S.~D.}\ \bibnamefont
  {Huber}},\ }\href@noop {} {\bibfield  {journal} {\bibinfo  {journal} {Nat.
  Phys.}\ }\textbf {\bibinfo {volume} {12}},\ \bibinfo {pages} {621} (\bibinfo
  {year} {2016})}\BibitemShut {NoStop}%
\bibitem [{\citenamefont {Jotzu}\ \emph {et~al.}(2014)\citenamefont {Jotzu},
  \citenamefont {Messer}, \citenamefont {Desbuquois}, \citenamefont {Lebrat},
  \citenamefont {Uehlinger}, \citenamefont {Greif},\ and\ \citenamefont
  {Esslinger}}]{Jotzu2014NatLett}%
  \BibitemOpen
  \bibfield  {author} {\bibinfo {author} {\bibfnamefont {G.}~\bibnamefont
  {Jotzu}}, \bibinfo {author} {\bibfnamefont {M.}~\bibnamefont {Messer}},
  \bibinfo {author} {\bibfnamefont {R.}~\bibnamefont {Desbuquois}}, \bibinfo
  {author} {\bibfnamefont {M.}~\bibnamefont {Lebrat}}, \bibinfo {author}
  {\bibfnamefont {T.}~\bibnamefont {Uehlinger}}, \bibinfo {author}
  {\bibfnamefont {D.}~\bibnamefont {Greif}}, \ and\ \bibinfo {author}
  {\bibfnamefont {T.}~\bibnamefont {Esslinger}},\ }\href@noop {} {\bibfield
  {journal} {\bibinfo  {journal} {Nature}\ }\textbf {\bibinfo {volume} {515}},\
  \bibinfo {pages} {237} (\bibinfo {year} {2014})}\BibitemShut {NoStop}%
\bibitem [{\citenamefont {Kane}\ and\ \citenamefont
  {Lubensky}(2014)}]{Kane2014NatPhys}%
  \BibitemOpen
  \bibfield  {author} {\bibinfo {author} {\bibfnamefont {C.~L.}\ \bibnamefont
  {Kane}}\ and\ \bibinfo {author} {\bibfnamefont {T.~C.}\ \bibnamefont
  {Lubensky}},\ }\href@noop {} {\bibfield  {journal} {\bibinfo  {journal} {Nat.
  Phys.}\ }\textbf {\bibinfo {volume} {10}},\ \bibinfo {pages} {39} (\bibinfo
  {year} {2014})}\BibitemShut {NoStop}%
\bibitem [{\citenamefont {Lu}\ \emph {et~al.}(2015)\citenamefont {Lu},
  \citenamefont {Wang}, \citenamefont {Ye}, \citenamefont {Ran}, \citenamefont
  {Fu}, \citenamefont {Joannopoulos},\ and\ \citenamefont
  {Solja{\v{c}}i{\'c}}}]{Lu2015Sci}%
  \BibitemOpen
  \bibfield  {author} {\bibinfo {author} {\bibfnamefont {L.}~\bibnamefont
  {Lu}}, \bibinfo {author} {\bibfnamefont {Z.}~\bibnamefont {Wang}}, \bibinfo
  {author} {\bibfnamefont {D.}~\bibnamefont {Ye}}, \bibinfo {author}
  {\bibfnamefont {L.}~\bibnamefont {Ran}}, \bibinfo {author} {\bibfnamefont
  {L.}~\bibnamefont {Fu}}, \bibinfo {author} {\bibfnamefont {J.~D.}\
  \bibnamefont {Joannopoulos}}, \ and\ \bibinfo {author} {\bibfnamefont
  {M.}~\bibnamefont {Solja{\v{c}}i{\'c}}},\ }\href@noop {} {\bibfield
  {journal} {\bibinfo  {journal} {Science}\ }\textbf {\bibinfo {volume}
  {349}},\ \bibinfo {pages} {622} (\bibinfo {year} {2015})}\BibitemShut
  {NoStop}%
\bibitem [{\citenamefont {Lu}\ \emph {et~al.}(2017)\citenamefont {Lu},
  \citenamefont {Qiu}, \citenamefont {Ye}, \citenamefont {Fan}, \citenamefont
  {Ke}, \citenamefont {Zhang},\ and\ \citenamefont {Liu}}]{Lu2017NatPhys}%
  \BibitemOpen
  \bibfield  {author} {\bibinfo {author} {\bibfnamefont {J.}~\bibnamefont
  {Lu}}, \bibinfo {author} {\bibfnamefont {C.}~\bibnamefont {Qiu}}, \bibinfo
  {author} {\bibfnamefont {L.}~\bibnamefont {Ye}}, \bibinfo {author}
  {\bibfnamefont {X.}~\bibnamefont {Fan}}, \bibinfo {author} {\bibfnamefont
  {M.}~\bibnamefont {Ke}}, \bibinfo {author} {\bibfnamefont {F.}~\bibnamefont
  {Zhang}}, \ and\ \bibinfo {author} {\bibfnamefont {Z.}~\bibnamefont {Liu}},\
  }\href@noop {} {\bibfield  {journal} {\bibinfo  {journal} {Nat. Phys.}\
  }\textbf {\bibinfo {volume} {13}},\ \bibinfo {pages} {369} (\bibinfo {year}
  {2017})}\BibitemShut {NoStop}%
\bibitem [{\citenamefont {Meier}\ \emph {et~al.}(2016)\citenamefont {Meier},
  \citenamefont {An},\ and\ \citenamefont {Gadway}}]{Meier2016NatComm}%
  \BibitemOpen
  \bibfield  {author} {\bibinfo {author} {\bibfnamefont {E.~J.}\ \bibnamefont
  {Meier}}, \bibinfo {author} {\bibfnamefont {F.~A.}\ \bibnamefont {An}}, \
  and\ \bibinfo {author} {\bibfnamefont {B.}~\bibnamefont {Gadway}},\
  }\href@noop {} {\bibfield  {journal} {\bibinfo  {journal} {Nat. Commun.}\
  }\textbf {\bibinfo {volume} {7}},\ \bibinfo {pages} {13986} (\bibinfo {year}
  {2016})}\BibitemShut {NoStop}%
\bibitem [{\citenamefont {Bliokh}\ \emph {et~al.}(2015)\citenamefont {Bliokh},
  \citenamefont {Smirnova},\ and\ \citenamefont {Nori}}]{Bliokh2015Sci}%
  \BibitemOpen
  \bibfield  {author} {\bibinfo {author} {\bibfnamefont {K.~Y.}\ \bibnamefont
  {Bliokh}}, \bibinfo {author} {\bibfnamefont {D.}~\bibnamefont {Smirnova}}, \
  and\ \bibinfo {author} {\bibfnamefont {F.}~\bibnamefont {Nori}},\ }\href@noop
  {} {\bibfield  {journal} {\bibinfo  {journal} {Science}\ }\textbf {\bibinfo
  {volume} {348}},\ \bibinfo {pages} {1448} (\bibinfo {year}
  {2015})}\BibitemShut {NoStop}%
\bibitem [{\citenamefont {Klembt}\ \emph {et~al.}(2018)\citenamefont {Klembt},
  \citenamefont {Harder}, \citenamefont {Egorov}, \citenamefont {Winkler},
  \citenamefont {Ge}, \citenamefont {Bandres}, \citenamefont {Emmerling},
  \citenamefont {Worschech}, \citenamefont {Liew}, \citenamefont {Segev},
  \citenamefont {Schneider},\ and\ \citenamefont
  {H{\"o}fling}}]{Klembt2018Nat}%
  \BibitemOpen
  \bibfield  {author} {\bibinfo {author} {\bibfnamefont {S.}~\bibnamefont
  {Klembt}}, \bibinfo {author} {\bibfnamefont {T.~H.}\ \bibnamefont {Harder}},
  \bibinfo {author} {\bibfnamefont {O.~A.}\ \bibnamefont {Egorov}}, \bibinfo
  {author} {\bibfnamefont {K.}~\bibnamefont {Winkler}}, \bibinfo {author}
  {\bibfnamefont {R.}~\bibnamefont {Ge}}, \bibinfo {author} {\bibfnamefont
  {M.~A.}\ \bibnamefont {Bandres}}, \bibinfo {author} {\bibfnamefont
  {M.}~\bibnamefont {Emmerling}}, \bibinfo {author} {\bibfnamefont
  {L.}~\bibnamefont {Worschech}}, \bibinfo {author} {\bibfnamefont {T.~C.~H.}\
  \bibnamefont {Liew}}, \bibinfo {author} {\bibfnamefont {M.}~\bibnamefont
  {Segev}}, \bibinfo {author} {\bibfnamefont {C.}~\bibnamefont {Schneider}}, \
  and\ \bibinfo {author} {\bibfnamefont {S.}~\bibnamefont {H{\"o}fling}},\
  }\href@noop {} {\bibfield  {journal} {\bibinfo  {journal} {Nature}\ }\textbf
  {\bibinfo {volume} {562}},\ \bibinfo {pages} {552} (\bibinfo {year}
  {2018})}\BibitemShut {NoStop}%
\bibitem [{\citenamefont {Klembt}\ \emph {et~al.}(2017)\citenamefont {Klembt},
  \citenamefont {Harder}, \citenamefont {Egorov}, \citenamefont {Winkler},
  \citenamefont {Suchomel}, \citenamefont {Beierlein}, \citenamefont
  {Emmerling}, \citenamefont {Schneider},\ and\ \citenamefont
  {H{\"o}fling}}]{Klembt2017APL}%
  \BibitemOpen
  \bibfield  {author} {\bibinfo {author} {\bibfnamefont {S.}~\bibnamefont
  {Klembt}}, \bibinfo {author} {\bibfnamefont {T.~H.}\ \bibnamefont {Harder}},
  \bibinfo {author} {\bibfnamefont {O.~A.}\ \bibnamefont {Egorov}}, \bibinfo
  {author} {\bibfnamefont {K.}~\bibnamefont {Winkler}}, \bibinfo {author}
  {\bibfnamefont {H.}~\bibnamefont {Suchomel}}, \bibinfo {author}
  {\bibfnamefont {J.}~\bibnamefont {Beierlein}}, \bibinfo {author}
  {\bibfnamefont {M.}~\bibnamefont {Emmerling}}, \bibinfo {author}
  {\bibfnamefont {C.}~\bibnamefont {Schneider}}, \ and\ \bibinfo {author}
  {\bibfnamefont {S.}~\bibnamefont {H{\"o}fling}},\ }\href@noop {} {\bibfield
  {journal} {\bibinfo  {journal} {Appl. Phys. Lett.}\ }\textbf {\bibinfo
  {volume} {111}},\ \bibinfo {pages} {231102} (\bibinfo {year}
  {2017})}\BibitemShut {NoStop}%
\bibitem [{\citenamefont {Haldane}(1988)}]{Haldane1988PRL}%
  \BibitemOpen
  \bibfield  {author} {\bibinfo {author} {\bibfnamefont {F.~D.~M.}\
  \bibnamefont {Haldane}},\ }\href@noop {} {\bibfield  {journal} {\bibinfo
  {journal} {Phys. Rev. Lett.}\ }\textbf {\bibinfo {volume} {61}},\ \bibinfo
  {pages} {2015} (\bibinfo {year} {1988})}\BibitemShut {NoStop}%
\bibitem [{\citenamefont {Qian}\ \emph {et~al.}()\citenamefont {Qian},
  \citenamefont {Apigo}, \citenamefont {Prodan}, \citenamefont {Barlas},\ and\
  \citenamefont {Prodan}}]{Kai2018arXiv}%
  \BibitemOpen
  \bibfield  {author} {\bibinfo {author} {\bibfnamefont {K.}~\bibnamefont
  {Qian}}, \bibinfo {author} {\bibfnamefont {D.~J.}\ \bibnamefont {Apigo}},
  \bibinfo {author} {\bibfnamefont {C.}~\bibnamefont {Prodan}}, \bibinfo
  {author} {\bibfnamefont {Y.}~\bibnamefont {Barlas}}, \ and\ \bibinfo {author}
  {\bibfnamefont {E.}~\bibnamefont {Prodan}},\ }\href@noop {} {\bibinfo
  {journal} {arXiv:1803.08781}\ }\BibitemShut {NoStop}%
\bibitem [{\citenamefont {Vila}\ \emph {et~al.}(2017)\citenamefont {Vila},
  \citenamefont {Pal},\ and\ \citenamefont {Ruzzene}}]{Vila2017PRB}%
  \BibitemOpen
\bibfield  {journal} {  }\bibfield  {author} {\bibinfo {author} {\bibfnamefont
  {J.}~\bibnamefont {Vila}}, \bibinfo {author} {\bibfnamefont {R.~K.}\
  \bibnamefont {Pal}}, \ and\ \bibinfo {author} {\bibfnamefont
  {M.}~\bibnamefont {Ruzzene}},\ }\href@noop {} {\bibfield  {journal} {\bibinfo
   {journal} {Phys. Rev. B}\ }\textbf {\bibinfo {volume} {96}},\ \bibinfo
  {pages} {134307} (\bibinfo {year} {2017})}\BibitemShut {NoStop}%
\bibitem [{\citenamefont {Pal}\ and\ \citenamefont
  {Ruzzene}(2017)}]{Pal2017NJP}%
  \BibitemOpen
  \bibfield  {author} {\bibinfo {author} {\bibfnamefont {R.~K.}\ \bibnamefont
  {Pal}}\ and\ \bibinfo {author} {\bibfnamefont {M.}~\bibnamefont {Ruzzene}},\
  }\href@noop {} {\bibfield  {journal} {\bibinfo  {journal} {New J. Phys.}\
  }\textbf {\bibinfo {volume} {19}},\ \bibinfo {pages} {025001} (\bibinfo
  {year} {2017})}\BibitemShut {NoStop}%
\bibitem [{\citenamefont {Asb{\'o}th}\ \emph {et~al.}(2015)\citenamefont
  {Asb{\'o}th}, \citenamefont {Oroszl{\'a}ny},\ and\ \citenamefont
  {P{\'a}lyi}}]{ShortCourse}%
  \BibitemOpen
  \bibfield  {author} {\bibinfo {author} {\bibfnamefont {J.~K.}\ \bibnamefont
  {Asb{\'o}th}}, \bibinfo {author} {\bibfnamefont {L.}~\bibnamefont
  {Oroszl{\'a}ny}}, \ and\ \bibinfo {author} {\bibfnamefont {A.}~\bibnamefont
  {P{\'a}lyi}},\ }\href@noop {} {\emph {\bibinfo {title} {A Short Course on
  Topological Insulators}}}\ (\bibinfo  {publisher} {Springer, New York},\
  \bibinfo {year} {2015})\BibitemShut {NoStop}%
\bibitem [{\citenamefont {Li}\ \emph {et~al.}(2014)\citenamefont {Li},
  \citenamefont {Xu},\ and\ \citenamefont {Chen}}]{ChenPRB2014}%
  \BibitemOpen
  \bibfield  {author} {\bibinfo {author} {\bibfnamefont {L.}~\bibnamefont
  {Li}}, \bibinfo {author} {\bibfnamefont {Z.}~\bibnamefont {Xu}}, \ and\
  \bibinfo {author} {\bibfnamefont {S.}~\bibnamefont {Chen}},\ }\href@noop {}
  {\bibfield  {journal} {\bibinfo  {journal} {Phys. Rev. B}\ }\textbf {\bibinfo
  {volume} {89}},\ \bibinfo {pages} {085111} (\bibinfo {year}
  {2014})}\BibitemShut {NoStop}%
\bibitem [{\citenamefont {Delplace}\ \emph {et~al.}(2011)\citenamefont
  {Delplace}, \citenamefont {Ullmo},\ and\ \citenamefont
  {Montambaux}}]{Delplace11PRB}%
  \BibitemOpen
  \bibfield  {author} {\bibinfo {author} {\bibfnamefont {P.}~\bibnamefont
  {Delplace}}, \bibinfo {author} {\bibfnamefont {D.}~\bibnamefont {Ullmo}}, \
  and\ \bibinfo {author} {\bibfnamefont {G.}~\bibnamefont {Montambaux}},\
  }\href@noop {} {\bibfield  {journal} {\bibinfo  {journal} {Phys. Rev. B}\
  }\textbf {\bibinfo {volume} {84}},\ \bibinfo {pages} {195452} (\bibinfo
  {year} {2011})}\BibitemShut {NoStop}%
\bibitem [{\citenamefont {Ahn}\ \emph {et~al.}(2004)\citenamefont {Ahn},
  \citenamefont {Lookman},\ and\ \citenamefont {Bishop}}]{AhnNature2004}%
  \BibitemOpen
  \bibfield  {author} {\bibinfo {author} {\bibfnamefont {K.~H.}\ \bibnamefont
  {Ahn}}, \bibinfo {author} {\bibfnamefont {T.}~\bibnamefont {Lookman}}, \ and\
  \bibinfo {author} {\bibfnamefont {A.~R.}\ \bibnamefont {Bishop}},\
  }\href@noop {} {\bibfield  {journal} {\bibinfo  {journal} {Nature (London)}\
  }\textbf {\bibinfo {volume} {428}},\ \bibinfo {pages} {401} (\bibinfo {year}
  {2004})}\BibitemShut {NoStop}%
\bibitem [{\citenamefont {Barsch}\ and\ \citenamefont
  {Krumhansl}(1984)}]{Barsch1984PRL}%
  \BibitemOpen
  \bibfield  {author} {\bibinfo {author} {\bibfnamefont {G.~R.}\ \bibnamefont
  {Barsch}}\ and\ \bibinfo {author} {\bibfnamefont {J.~A.}\ \bibnamefont
  {Krumhansl}},\ }\href@noop {} {\bibfield  {journal} {\bibinfo  {journal}
  {Phys. Rev. Lett.}\ }\textbf {\bibinfo {volume} {53}},\ \bibinfo {pages}
  {1069} (\bibinfo {year} {1984})}\BibitemShut {NoStop}%
\bibitem [{\citenamefont {Mesaros}\ \emph {et~al.}(2010)\citenamefont
  {Mesaros}, \citenamefont {Papanikolaou}, \citenamefont {Flipse},
  \citenamefont {Sadri},\ and\ \citenamefont {Zaanen}}]{Mesaros2010PRB}%
  \BibitemOpen
  \bibfield  {author} {\bibinfo {author} {\bibfnamefont {A.}~\bibnamefont
  {Mesaros}}, \bibinfo {author} {\bibfnamefont {S.}~\bibnamefont
  {Papanikolaou}}, \bibinfo {author} {\bibfnamefont {C.~F.~J.}\ \bibnamefont
  {Flipse}}, \bibinfo {author} {\bibfnamefont {D.}~\bibnamefont {Sadri}}, \
  and\ \bibinfo {author} {\bibfnamefont {J.}~\bibnamefont {Zaanen}},\
  }\href@noop {} {\bibfield  {journal} {\bibinfo  {journal} {Phys. Rev. B}\
  }\textbf {\bibinfo {volume} {82}},\ \bibinfo {pages} {205119} (\bibinfo
  {year} {2010})}\BibitemShut {NoStop}%
\bibitem [{\citenamefont {Takahashi}\ and\ \citenamefont
  {Murakami}(2011)}]{Takahashi2011PRL}%
  \BibitemOpen
  \bibfield  {author} {\bibinfo {author} {\bibfnamefont {R.}~\bibnamefont
  {Takahashi}}\ and\ \bibinfo {author} {\bibfnamefont {S.}~\bibnamefont
  {Murakami}},\ }\href@noop {} {\bibfield  {journal} {\bibinfo  {journal}
  {Phys. Rev. Lett.}\ }\textbf {\bibinfo {volume} {107}},\ \bibinfo {pages}
  {166805} (\bibinfo {year} {2011})}\BibitemShut {NoStop}%
\bibitem [{\citenamefont {Phillips}\ and\ \citenamefont
  {Mele}(2015)}]{Phillips2015PRB}%
  \BibitemOpen
  \bibfield  {author} {\bibinfo {author} {\bibfnamefont {M.}~\bibnamefont
  {Phillips}}\ and\ \bibinfo {author} {\bibfnamefont {E.~J.}\ \bibnamefont
  {Mele}},\ }\href@noop {} {\bibfield  {journal} {\bibinfo  {journal} {Phys.
  Rev. B}\ }\textbf {\bibinfo {volume} {91}},\ \bibinfo {pages} {125404}
  (\bibinfo {year} {2015})}\BibitemShut {NoStop}%
\bibitem [{\citenamefont {Slager}\ \emph {et~al.}(2016)\citenamefont {Slager},
  \citenamefont {Juri{\v{c}}i{\'c}}, \citenamefont {Lahtinen},\ and\
  \citenamefont {Zaanen}}]{Slager2016PRB}%
  \BibitemOpen
  \bibfield  {author} {\bibinfo {author} {\bibfnamefont {R.-J.}\ \bibnamefont
  {Slager}}, \bibinfo {author} {\bibfnamefont {V.}~\bibnamefont
  {Juri{\v{c}}i{\'c}}}, \bibinfo {author} {\bibfnamefont {V.}~\bibnamefont
  {Lahtinen}}, \ and\ \bibinfo {author} {\bibfnamefont {J.}~\bibnamefont
  {Zaanen}},\ }\href@noop {} {\bibfield  {journal} {\bibinfo  {journal} {Phys.
  Rev. B}\ }\textbf {\bibinfo {volume} {93}},\ \bibinfo {pages} {245406}
  (\bibinfo {year} {2016})}\BibitemShut {NoStop}%
\bibitem [{\citenamefont {Rhim}\ \emph {et~al.}(2018)\citenamefont {Rhim},
  \citenamefont {Bardarson},\ and\ \citenamefont {Slager}}]{Rhim2018PRB}%
  \BibitemOpen
  \bibfield  {author} {\bibinfo {author} {\bibfnamefont {J.-W.}\ \bibnamefont
  {Rhim}}, \bibinfo {author} {\bibfnamefont {J.~H.}\ \bibnamefont {Bardarson}},
  \ and\ \bibinfo {author} {\bibfnamefont {R.-J.}\ \bibnamefont {Slager}},\
  }\href@noop {} {\bibfield  {journal} {\bibinfo  {journal} {Phys. Rev. B}\
  }\textbf {\bibinfo {volume} {97}},\ \bibinfo {pages} {115143} (\bibinfo
  {year} {2018})}\BibitemShut {NoStop}%
\bibitem [{Foo({\natexlab{a}})}]{FootNote2}%
  \BibitemOpen
  \href@noop {} {\emph {\bibinfo {title} {\rm Topological analysis used for
  zero-angle grain boundaries in graphene in Ref.~\onlinecite{Phillips2015PRB}
  could be also applied to the TB and APB studied here, particularly with
  regards to mirror-odd states. It is noteworthy that flat bands occur only
  within a part of the projected first Brillouin zone for zero-angle grain
  boundaries or OE in graphenes, but occur throughout the entire projected zone
  for the TB/APB/OE for the lattice in Fig.~\ref{fig:model}.}}}\BibitemShut
  {Stop}%
\bibitem [{\citenamefont {Schnyder}\ \emph {et~al.}(2008)\citenamefont
  {Schnyder}, \citenamefont {Ryu}, \citenamefont {Furusaki},\ and\
  \citenamefont {Ludwig}}]{Schnyder2008PRB}%
  \BibitemOpen
  \bibfield  {author} {\bibinfo {author} {\bibfnamefont {A.~P.}\ \bibnamefont
  {Schnyder}}, \bibinfo {author} {\bibfnamefont {S.}~\bibnamefont {Ryu}},
  \bibinfo {author} {\bibfnamefont {A.}~\bibnamefont {Furusaki}}, \ and\
  \bibinfo {author} {\bibfnamefont {A.~W.~W.}\ \bibnamefont {Ludwig}},\
  }\href@noop {} {\bibfield  {journal} {\bibinfo  {journal} {Phys. Rev. B}\
  }\textbf {\bibinfo {volume} {78}},\ \bibinfo {pages} {195125} (\bibinfo
  {year} {2008})}\BibitemShut {NoStop}%
\bibitem [{\citenamefont {Qi}\ \emph {et~al.}(2008)\citenamefont {Qi},
  \citenamefont {Hughes},\ and\ \citenamefont {Zhang}}]{Qi2008PRB}%
  \BibitemOpen
  \bibfield  {author} {\bibinfo {author} {\bibfnamefont {X.-L.}\ \bibnamefont
  {Qi}}, \bibinfo {author} {\bibfnamefont {T.~L.}\ \bibnamefont {Hughes}}, \
  and\ \bibinfo {author} {\bibfnamefont {S.-C.}\ \bibnamefont {Zhang}},\
  }\href@noop {} {\bibfield  {journal} {\bibinfo  {journal} {Phys. Rev. B}\
  }\textbf {\bibinfo {volume} {78}},\ \bibinfo {pages} {195424} (\bibinfo
  {year} {2008})}\BibitemShut {NoStop}%
\bibitem [{\citenamefont {Kitaev}(2009)}]{Kitaev2009AIP}%
  \BibitemOpen
  \bibfield  {author} {\bibinfo {author} {\bibfnamefont {A.}~\bibnamefont
  {Kitaev}},\ }\href@noop {} {\bibfield  {journal} {\bibinfo  {journal} {AIP
  Conf. Proc.}\ }\textbf {\bibinfo {volume} {1134}},\ \bibinfo {pages} {22}
  (\bibinfo {year} {2009})}\BibitemShut {NoStop}%
\bibitem [{\citenamefont {Ryu}\ \emph {et~al.}(2010)\citenamefont {Ryu},
  \citenamefont {Schnyder}, \citenamefont {Furusaki},\ and\ \citenamefont
  {Ludwig}}]{Ryu2010tNJP}%
  \BibitemOpen
  \bibfield  {author} {\bibinfo {author} {\bibfnamefont {S.}~\bibnamefont
  {Ryu}}, \bibinfo {author} {\bibfnamefont {A.~P.}\ \bibnamefont {Schnyder}},
  \bibinfo {author} {\bibfnamefont {A.}~\bibnamefont {Furusaki}}, \ and\
  \bibinfo {author} {\bibfnamefont {A.~W.~W.}\ \bibnamefont {Ludwig}},\
  }\href@noop {} {\bibfield  {journal} {\bibinfo  {journal} {New J. Phys.}\
  }\textbf {\bibinfo {volume} {12}},\ \bibinfo {pages} {065010} (\bibinfo
  {year} {2010})}\BibitemShut {NoStop}%
\bibitem [{\citenamefont {Prodan}\ and\ \citenamefont
  {Schulz-Baldes}(2016)}]{ProdanBook2016}%
  \BibitemOpen
  \bibfield  {author} {\bibinfo {author} {\bibfnamefont {E.}~\bibnamefont
  {Prodan}}\ and\ \bibinfo {author} {\bibfnamefont {H.}~\bibnamefont
  {Schulz-Baldes}},\ }\href@noop {} {\emph {\bibinfo {title} {Bulk and Boundary
  Invariants for Complex Topological Insulators: From $K$-Theory to Physics}}}\
  (\bibinfo  {publisher} {Springer, New York},\ \bibinfo {year}
  {2016})\BibitemShut {NoStop}%
\bibitem [{\citenamefont {Zak}(1989)}]{Zak1989PRL}%
  \BibitemOpen
  \bibfield  {author} {\bibinfo {author} {\bibfnamefont {J.}~\bibnamefont
  {Zak}},\ }\href@noop {} {\bibfield  {journal} {\bibinfo  {journal} {Phys.
  Rev. Lett.}\ }\textbf {\bibinfo {volume} {62}},\ \bibinfo {pages} {2747}
  (\bibinfo {year} {1989})}\BibitemShut {NoStop}%
\bibitem [{Foo({\natexlab{b}})}]{FootNote1}%
  \BibitemOpen
  \href@noop {} {\emph {\bibinfo {title} {\rm See Supplemental Material at [~]
  for lattice energy used to obtain distorted lattice
  configurations.}}}\BibitemShut {Stop}%
\bibitem [{\citenamefont {Shenoy}\ \emph {et~al.}(1999)\citenamefont {Shenoy},
  \citenamefont {Lookman}, \citenamefont {Saxena},\ and\ \citenamefont
  {Bishop}}]{Shenoy1999PRB}%
  \BibitemOpen
  \bibfield  {author} {\bibinfo {author} {\bibfnamefont {S.~R.}\ \bibnamefont
  {Shenoy}}, \bibinfo {author} {\bibfnamefont {T.}~\bibnamefont {Lookman}},
  \bibinfo {author} {\bibfnamefont {A.}~\bibnamefont {Saxena}}, \ and\ \bibinfo
  {author} {\bibfnamefont {A.~R.}\ \bibnamefont {Bishop}},\ }\href@noop {}
  {\bibfield  {journal} {\bibinfo  {journal} {Phys. Rev. B}\ }\textbf {\bibinfo
  {volume} {60}},\ \bibinfo {pages} {R12537} (\bibinfo {year}
  {1999})}\BibitemShut {NoStop}%
\bibitem [{\citenamefont {Kim}\ \emph {et~al.}(2016)\citenamefont {Kim},
  \citenamefont {Lee}, \citenamefont {Kim}, \citenamefont {Choi}, \citenamefont
  {Chang}, \citenamefont {Won}, \citenamefont {Kwon}, \citenamefont {Kim},
  \citenamefont {Hyun}, \citenamefont {Kim}, \citenamefont {Koo}, \citenamefont
  {Choi}, \citenamefont {Kim},\ and\ \citenamefont {Baek}}]{Kim2016NatComm}%
  \BibitemOpen
  \bibfield  {author} {\bibinfo {author} {\bibfnamefont {K.-C.}\ \bibnamefont
  {Kim}}, \bibinfo {author} {\bibfnamefont {J.}~\bibnamefont {Lee}}, \bibinfo
  {author} {\bibfnamefont {B.~K.}\ \bibnamefont {Kim}}, \bibinfo {author}
  {\bibfnamefont {W.~Y.}\ \bibnamefont {Choi}}, \bibinfo {author}
  {\bibfnamefont {H.~J.}\ \bibnamefont {Chang}}, \bibinfo {author}
  {\bibfnamefont {S.~O.}\ \bibnamefont {Won}}, \bibinfo {author} {\bibfnamefont
  {B.}~\bibnamefont {Kwon}}, \bibinfo {author} {\bibfnamefont {S.~K.}\
  \bibnamefont {Kim}}, \bibinfo {author} {\bibfnamefont {D.-B.}\ \bibnamefont
  {Hyun}}, \bibinfo {author} {\bibfnamefont {H.~J.}\ \bibnamefont {Kim}},
  \bibinfo {author} {\bibfnamefont {H.~C.}\ \bibnamefont {Koo}}, \bibinfo
  {author} {\bibfnamefont {J.-H.}\ \bibnamefont {Choi}}, \bibinfo {author}
  {\bibfnamefont {J.-S.}\ \bibnamefont {Kim}}, \ and\ \bibinfo {author}
  {\bibfnamefont {S.-H.}\ \bibnamefont {Baek}},\ }\href@noop {} {\bibfield
  {journal} {\bibinfo  {journal} {Nat. Commun.}\ }\textbf {\bibinfo {volume}
  {7}},\ \bibinfo {pages} {12449} (\bibinfo {year} {2016})}\BibitemShut
  {NoStop}%
\bibitem [{Foo({\natexlab{c}})}]{FootNote3}%
  \BibitemOpen
  \href@noop {} {\emph {\bibinfo {title} {\rm Even in the presence of
  disorders, the winding number, expressed in real space, would be unchanged as
  far as the chiral symmetry is preserved and the bulk spectral gap remains
  open~\cite{ProdanBook2016}. The bulk-boundary correspondence would still
  apply to such cases and the zero-energy states found at TB/APB/OE in the
  current work would be topologically protected.}}}\BibitemShut {Stop}%
\end{thebibliography}%


\begin{thebibliography}{4}
\expandafter\ifx\csname natexlab\endcsname\relax\def\natexlab#1{#1}\fi
\expandafter\ifx\csname bibnamefont\endcsname\relax
  \def\bibnamefont#1{#1}\fi
\expandafter\ifx\csname bibfnamefont\endcsname\relax
  \def\bibfnamefont#1{#1}\fi
\expandafter\ifx\csname citenamefont\endcsname\relax
  \def\citenamefont#1{#1}\fi
\expandafter\ifx\csname url\endcsname\relax
  \def\url#1{\texttt{#1}}\fi
\expandafter\ifx\csname urlprefix\endcsname\relax\def\urlprefix{URL }\fi
\providecommand{\bibinfo}[2]{#2}
\providecommand{\eprint}[2][]{\url{#2}}

\bibitem[{\citenamefont{Ahn et~al.}(2003)\citenamefont{Ahn, Lookman, Saxena,
  and Bishop}}]{AhnPRB2003}
\bibinfo{author}{\bibfnamefont{K.~H.} \bibnamefont{Ahn}},
  \bibinfo{author}{\bibfnamefont{T.}~\bibnamefont{Lookman}},
  \bibinfo{author}{\bibfnamefont{A.}~\bibnamefont{Saxena}}, \bibnamefont{and}
  \bibinfo{author}{\bibfnamefont{A.~R.} \bibnamefont{Bishop}},
  \bibinfo{journal}{Phys. Rev. B} \textbf{\bibinfo{volume}{68}},
  \bibinfo{pages}{092101} (\bibinfo{year}{2003}).

\bibitem[{\citenamefont{Ahn et~al.}(2004)\citenamefont{Ahn, Lookman, and
  Bishop}}]{AhnNature2004}
\bibinfo{author}{\bibfnamefont{K.~H.} \bibnamefont{Ahn}},
  \bibinfo{author}{\bibfnamefont{T.}~\bibnamefont{Lookman}}, \bibnamefont{and}
  \bibinfo{author}{\bibfnamefont{A.~R.} \bibnamefont{Bishop}},
  \bibinfo{journal}{Nature (London)} \textbf{\bibinfo{volume}{428}},
  \bibinfo{pages}{401} (\bibinfo{year}{2004}).

\bibitem[{\citenamefont{Ahn et~al.}(2005)\citenamefont{Ahn, Lookman, Saxena,
  and Bishop}}]{AhnPRB2005}
\bibinfo{author}{\bibfnamefont{K.~H.} \bibnamefont{Ahn}},
  \bibinfo{author}{\bibfnamefont{T.}~\bibnamefont{Lookman}},
  \bibinfo{author}{\bibfnamefont{A.}~\bibnamefont{Saxena}}, \bibnamefont{and}
  \bibinfo{author}{\bibfnamefont{A.~R.} \bibnamefont{Bishop}},
  \bibinfo{journal}{Phys. Rev. B} \textbf{\bibinfo{volume}{71}},
  \bibinfo{pages}{212102} (\bibinfo{year}{2005}).

\bibitem[{\citenamefont{Ahn et~al.}(2013)\citenamefont{Ahn, Seman, Lookman, and
  Bishop}}]{AhnPRB2013}
\bibinfo{author}{\bibfnamefont{K.~H.} \bibnamefont{Ahn}},
  \bibinfo{author}{\bibfnamefont{T.~F.} \bibnamefont{Seman}},
  \bibinfo{author}{\bibfnamefont{T.}~\bibnamefont{Lookman}}, \bibnamefont{and}
  \bibinfo{author}{\bibfnamefont{A.~R.} \bibnamefont{Bishop}},
  \bibinfo{journal}{Phys. Rev. B} \textbf{\bibinfo{volume}{88}},
  \bibinfo{pages}{144415} (\bibinfo{year}{2013}).

\end{thebibliography}

\end{document}


\title{Flat Energy Bands within Antiphase and Twin Boundaries and at Open Edge in Topological Materials \\
SUPPLEMENTAL MATERIAL}

\author{Linghua Zhu}
    \affiliation{Department of Physics, New Jersey Institute of Technology, Newark, New Jersey 07102, USA\\}
\author{Emil Prodan}
    \affiliation{Department of Physics, Yeshiva University, New York, New York 10016, USA\\}    
\author{Keun Hyuk Ahn}
    \affiliation{Department of Physics, New Jersey Institute of Technology, Newark, New Jersey 07102, USA\\}

\maketitle
\section{Lattice energy for relaxed structural textures}
\begin{figure}[b]
    \centering
    \includegraphics[width=0.76\hsize,clip]{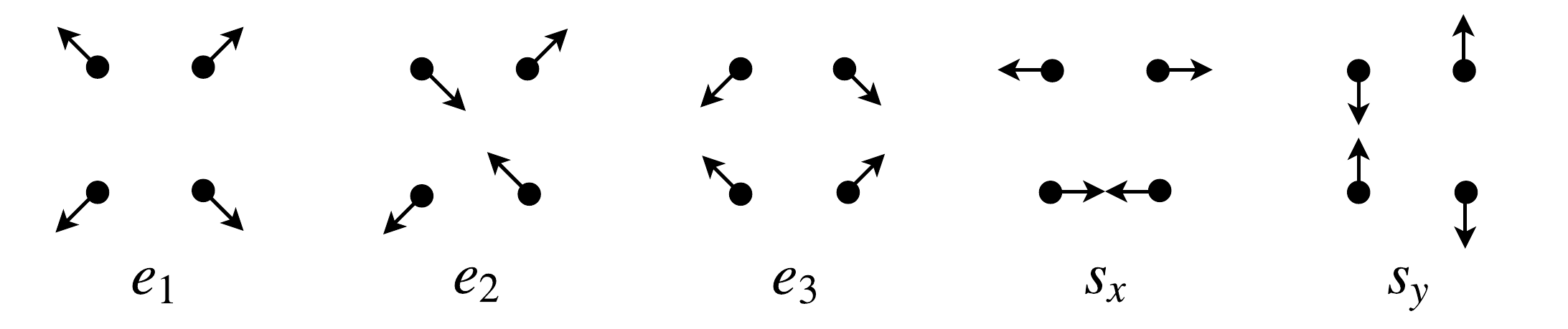}
    \caption{
     Atomic scale distortion modes for a square lattice~\cite{AhnPRB2003,AhnNature2004,AhnPRB2005,AhnPRB2013}. The modes $e_{1}$, $e_{2}$ and $e_{3}$ are long wavelength modes, and the modes $s_{x}$ and $s_{y}$ are short wavelength modes.
    } \label{fig:mode}
\end{figure}

Lattice distortion field should satisfy the compatibility conditions, that is, the bonds between atoms should not be broken or overlap with each other. 
Further, abrupt changes in lattice distortions usually cost a lot of lattice energy, and gradual changes of the lattice distortions would be favored energetically. 
To take both effects into account at the atomic scale, structural textures are obtained by relaxing an energy expression $E_{\rm lattice}$ written in terms of the atomic scale modes that represent the lattice distortions. Specifically, the distortion of a square lattice with a monatomic basis can be expressed in terms of five distortion modes, $e_{1}$, $e_{2}$, $e_{3}$, $s_{x}$, and $s_{y}$, shown in Fig.~\ref{fig:mode}, as discussed in detail in Refs.~\cite{AhnPRB2003,AhnNature2004,AhnPRB2005,AhnPRB2013}.
The energy expression $E_{\rm lattice}$, which gives rise to the ground states with $e>0,~d_{x}=0,~d_{y}\neq 0$ (equivalently with $e<0,~d_{x} \neq 0,~d_{y}=0$) illustrated in Fig.~\ref{fig:model} is as follow
\begin{eqnarray}
    &&E_{\rm lattice}=E_{s}+E_{l}+E_{c},  \\
    &&E_{s}=\sum_{\bf i}\left [ \frac{B}{2}(s_{x}^{2}+s_{y}^{2})+\frac{G_{1}}{4}(s_{x}^{4}+s_{y}^{4})+\frac{G_{2}}{2}s_{x}^{2}s_{y}^{2}) \right ]_{\bf i},  \\
    &&E_{l}=\sum_{\bf i}\left [ \frac{A_{1}}{2}e_{1}^{2}+\frac{A_{2}}{2}e_{2}^{2}+\frac{A_{3}}{2}e_{3}^{2} \right ]_{\bf i},  \\
    &&E_{c}=\sum_{\bf i}\left [ C_{3}(s_{x}^{2}-s_{y}^{2})e_{3} \right ]_{\bf i},
\end{eqnarray}
where $E_{s}$ is the energy for the short-wavelength modes including all symmertry-allowed terms up to the fourth order, $E_{l}$ the energy for the long-wavelength modes in the harmonic order, $E_{c}$ the coupling between the long- and short-wavelength modes, and ${\bf i}=\left [ i_{x},i_{y}\right ]$ designates coordinates before distortion.

For uniform phase, the energy minimization of $E_{\rm lattice}$ leads to $e_{1}=e_{2}=0$ and $e_{3}=-C_{3}(s^{2}_{x}-s^{2}_{y})/A_{3}$, and the minimized total energy per site is   
\begin{eqnarray}
    \frac{E_{\rm tot}^{\rm h,min}}{N^{2}} = \frac{B}{2}\left ( s^{2}_{x}+s^{2}_{y} \right )+\frac{1}{4}\left (G_{1}-\frac{2C_{3}^{2}}{A_{3}} \right) \left (s^{4}_{x}+s^{4}_{y} \right ) +\frac{1}{2} \left (G_{2}+\frac{2C_{3}^{2}}{A_{3}} \right)s_{x}^{2}s_{y}^{2}.
\end{eqnarray}
With $B<0$, $G_{1}-2C_{3}^{2}/A_{3}> 0$, $G_{2}+2C_{3}^{2}/A_{3}> 0 $, and $G_{2}+2C_{3}^{2}/A_{3}> G_{1}-2C_{3}^{2}/A_{3}$, the global energy minimum occurs for $s_{x}=\pm s_{\rm min}$, $s_{y}=0$ and equivalently for $s_{x}=0$, $s_{y}=\pm s^{\rm min}$, where $s^{\rm min}=\sqrt{BA_{3}/\left (  2C_{3}^{2}-G_{1}A_{3}\right )}$.
For the results in the main text, the parameter values are chosen as $A_{1}=7$, $A_{2}=4$, $A_{3}=6$, $B=-5$, $C_{3}=20$, $G_{1}=180$, and $G_{2}=100$, which satisfy the conditions specified above and give the minimum energy distortion $s^{\rm min}=0.327$ and $e_{3}^{\rm min}=0.357$, equivalently $e=0.252$ and $d_{x}=0.164,~d_{y}=0$ or $d_{x}=0,~d_{y}=0.164$ in Fig.~\ref{fig:model}.
For the inhomogenous configurations with TB and APB, the constraints among $e_{1}({\bf k}),~e_{2}({\bf k}),~e_{3}({\bf k}),~s_{x}({\bf k})$, and $s_{y}({\bf k})$ should be considered, as done in Refs.~\cite{AhnPRB2003,AhnNature2004,AhnPRB2005,AhnPRB2013}. With appropriately chosen initial configurations, the lattice configuration is relaxed using the Euler method applied to $E_{\rm lattice}$ to obtain the atomic scale profiles of TB, APB, and the zig-zag boundaries in Figs.~\ref{fig:TB}-\ref{fig:Mix} of the main text. 
\bibliography{SSH-Ref-Supp}